\documentclass[letterpaper]{sfchem}

\usepackage{graphicx}

\begin{document}

\title{ Carbon Chemistry in  interstellar clouds}

\author{Maryvonne Gerin\inst{1} 
\and David Foss\'e \inst{1} 
\and Evelyne Roueff \inst{2} }
   
\institute{LERMA, D\'epartement de  Physique de l'ENS et 
Observatoire de Paris, FRE2460 du CNRS,   24 Rue Lhomond, 
75231 Paris Cedex 05, France \and LUTH, Observatoire de Paris, 
FRE2462 du CNRS,   Place J. Janssen, 92195 Meudon cedex France }  

\authorrunning{Gerin, Foss\'e \& Roueff}


\titlerunning{Carbon Chemistry}

\maketitle 
\begin{abstract}
We discuss new developments of interstellar chemistry,
with particular emphasis on the carbon chemistry. We confirm 
that carbon chains and cycles are ubiquitous in the ISM and
closely chemically related to each other, and to carbon.
Investigation of the carbon budget in shielded and UV illuminated
gas shows that the inventory of interstellar molecules is
not complete and more complex molecules with 4 or more carbon atoms
must be present. Finally we discuss the consequences for
the evolution of clouds and conclude that the ubiquitous
presence of carbon chains and cycles is not  a necessary consequence of
a very young age for interstellar clouds.


\end{abstract}

\section{Introduction}
\label{intro}

 Nearly 120  molecules have now been  observed in various sources in
the interstellar medium. Emission or absorption lines from these
molecules carry important information on the physical conditions and
the chemical properties of the various sources where they have been 
recognized. However, only a very limited number of molecules is used
for that purpose, while much more information is available. The use 
of molecular lines detected with high resolution spectroscopy for 
studying the interstellar medium, and the star formation processes, 
are still largely unexplored. High resolution spectroscopy, either 
using heterodyne techniques in the radio domain, or with high
dispersive power spectrometers, yields not only the line intensities 
but information on the gas dynamics through the analysis of line 
profiles. Many lines 
can be analysed together for studying  physical conditions. 
Different approaches can be used to solve the radiative transfer
equation, from the simplest LTE model to sophisticated
multidimensional models. Also, other line properties can be exploited
to get information on the magnetic field, either directly from the
Zeeman effect on paramagnetic molecules, or indirectly with other
techniques (e.g.  \cite{Houde:03}). And finally, the molecular
abundances are the basic information for the chemistry. The question 
which arises then is how to select the best molecule and the best
lines to get useful  information. 

All interstellar molecules can not
be detected in the various environments of the ISM. This evolution of 
the emerging spectrum, hence of the chemical properties, is tightly 
connected to the physical conditions and environment of the source. 
With the advent of more sensitive receivers and better mapping
techniques, it is now possible to study extensively the chemical 
properties of specific sources. For example  \cite*{Tur:00} 
have performed detailed observational studies and modeling of a large 
set of interstellar molecules in cloud cores. We present in this paper 
new studies  of interstellar chemistry devoted to carbon chemistry 
and the condensation processes on dust grains, which make use of
molecular line observations.

 \section{Hydrocarbons in interstellar chemistry}
\label{Hydro}
Carbon is the fourth most abundant element in the interstellar medium, 
and also the more versatile for building molecules. Carbon chemistry 
can therefore be viewed as the core of interstellar chemistry. 
Most interstellar molecules (74 \%, or 85 out of 114) have at least 
one carbon atom. The heaviest and most complex molecules are organic 
molecules with carbon. This statistics does not take into account 
the PAHs, nor the DIBs carriers which most likely hypothesis are
 large organic molecules (\cite{Her:95}). Carbon, neutral or ionized, 
is also one of the main reactants in interstellar chemistry networks, due 
  to the large number of organic molecules, but also to  the 
reactivity and versatility of Carbon, which can participate in 
numerous chemical reactions at any temperature, from the very cold 
dense cores, to warm and hot gas. Therefore  understanding the carbon 
chemistry is of major importance in astrochemistry, and for star
formation. As a first step, we discuss  diffuse clouds and 
photon dominated regions,  and determine the carbon budget. 
We then discuss dark clouds chemistry and compare the carbon budget 
in all environments. At last we compare observations and
state of the art models. 

\subsection{Diffuse gas}
While CH, CH$^+$ and CN are known in the ISM since the 40's, the
detection of other molecules in diffuse gas has paused for a while. 
But recent deep observations have revealed and/or confirmed the
presence of diatomic, triatomic and even more complex molecules 
in diffuse gas. Carbon clusters (C$_2$, C$_3$) are now almost 
routinely detected in the visible towards bright stars 
(\cite{Maier:01}, \cite{Roueff:02}). C$_3$ can be seen  in the far 
infrared  through its low energy vibration mode 
( \cite{Giesen:01}, \cite{Cerni:00}); and maybe also C$_4$ 
\cite*{Cerni:02}. Detections in the far infrared are obtained
towards the massive star forming region SgrB2 close to the Galactic
Center, and sample more opaque clouds than typical diffuse clouds.
Typical numbers for column densities and abundance 
ratio are given in Table \ref{tab:dif}. 

Using radio telescopes  \cite*{Ll:00} have shown that CCH and
c-C$_3$H$_2$ are ubiquitous in diffuse gas, confirming previous 
work by \cite*{Cox:89}. These two species are spatially correlated  
and present a well defined abundance ratio 
N(CCH)/N(c-C$_3$H$_2$) = $21 \pm 6$ in diffuse clouds (\cite{Ll:00}). 
The previous work used strong background continuum sources such
as quasars to search for molecular absorption lines.
The supernova remnant, Cas~A, one of the strongest radio source 
in the sky, is a good target
to search for absorption lines.  \cite*{Bell:95} and \cite*{Bell:83} 
have reported the detection of C$_3$N and C$_4$H along this line of
sight. Though the physical conditions are not as accurately 
determined as in the diffuse clouds,  these observations confirm the 
presence of rather complex molecules in different environments exposed
to UV radiation.
\vskip 0.3cm
\begin{table}
\begin{center}
\caption{Typical column densities and abundances in diffuse clouds 
and in PDRs}
\renewcommand{\arraystretch}{1.2}
{\bf Diffuse clouds ; Absorption lines}\\
\begin{tabular}[h]{lcccc}
\hline
\hline
Molecule & N(X)  & ${\rm N(X)} \over \rm{N(H_2)}$ & 
$ {\rm N(C_2)} \over {\rm N(X)}$ & Ref.\\
 &  (cm$^{-2}$) & & & \\ 
\hline
C & $0.03 - 20 \times 10^{15}$ & $6.5 \pm 4 \times 10^{-6}$ & 0.008 &1
\\
CH & $1 - 10 \times 10^{13}$ & $5 \pm 2 \times 10^{-8}$ & 1 & 1 \\
C$_2$ &  $ 1 - 10\times 10^{13}$ & $5 \pm 2 \times 10^{-8}$ & 1 & 1 \\
C$_3$ &  $1.5 - 5 \times 10^{12}$ & $3 \times 10^{-9}$ & 15 & 2\\
C$_4$ ? &  $1.2 \times 10^{15} $ &  &  & 3 \\
\hline
 CCH & $0.17 - 3.5 \times 10^{13}$ & $3 \times 10^{-8}$ & 4 \\
c-C$_3$H$_2$ &  $0.5 - 1.7 \times 10^{12}$ & $1.5 \times 10^{-9}$ &
30& 4\\
C$_4$H & $ 0.9 - 2.5 \times 10^{13}$ & $1.5 \times 10^{-9}$ & 30 & 5
\\ 
\hline
\vspace{0.3em}
\end{tabular}
{\bf PDRs} \\
\begin{tabular}[h]{lcccc}
\hline
\hline
Molecule & N(X) & ${\rm N(X)} \over \rm{N(H_2)}$ & 
$ {\rm N(C_2H)} \over {\rm N(X)}$ & Ref.\\
 &  (cm$^{-2}$) & & & \\
\hline
CCH & $1.8 \times 10^{14}$ &  $1.7 \times 10^{-8}$ & 1 & 4\\
c-C$_3$H$_2$ & $1.0 \times 10^{13}$ & $1.1 \times 10^{-9}$ & 18 & 4 \\
C$_4$H & $2.0 \times 10^{13}$ & $2.0 \times 10^{-9}$ & 9 & 6\\
c-C$_3$H & $ 3.7\times 10^{12}$ & $3.7 \times 10^{-10}$ & 50 & 6\\
l-C$_3$H & $2.0 \times 10^{12}$ & $2.0 \times 10^{-10}$ & 90 & 6\\
l-C$_3$H$_2$ & $\leq 3.2 \times 10^{11}$ & & $\geq 500$  & 6 \\
\hline
\end{tabular}
\label{tab:dif}
\end{center}
{\small \it References : 1) \cite*{Fed:94}, \cite*{Rachford:02},
\cite*{Shuping:99} 2) \cite*{Roueff:02} 3) \cite*{Cerni:02} 
4) \cite*{Ll:00} 5)\cite*{Bell:83} 6) This work.}
\end{table}

\subsection{Photodissociation regions}
A natural question therefore arises : if carbon chains are present in
the diffuse interstellar medium, what happens in photon dissociation
regions ?
 As in diffuse clouds, the chemical processes are dominated by the 
radiation in PDRs, but the gas is denser. The radiation field, also 
more intense than in diffuse clouds, is generally well measured. Also, 
with the new data obtained with ISO, and particularly the mid-IR
emission due to PAHs, it is possible to estimate the gas density from 
the mid-IR data and compare with determination from molecular lines 
(see e.g. the case of the horsehead nebula (\cite{Abergel:03}, 
\cite{Teyssier:02}). We have detected  simple carbon chains and
cycles in four nearby photodissociation regions : the interface near 
HD~147889 in Ophiuchus, NGC~7023, IC~63 and the horsehead nebula. 
We report here the results for the last two sources but similar 
conclusions can be drawn for all PDRs (\cite{Fosse:03}). 
The horsehead nebula showed the brightest lines and deserved a more 
extensive study (\cite{Teyssier:02}).

\begin{figure}[h!]
\resizebox{\hsize}{!}
{\includegraphics{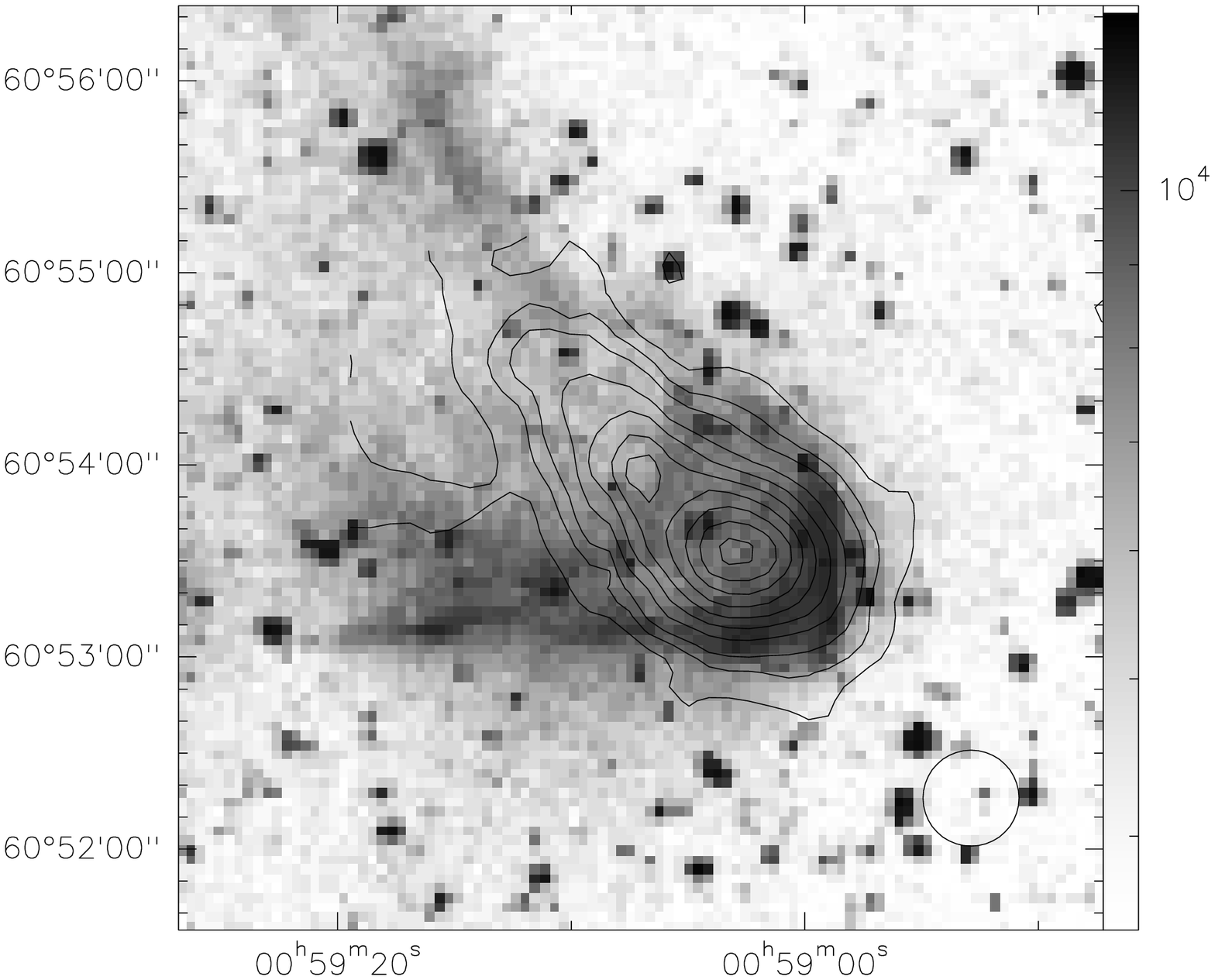} 
\includegraphics*[2cm,0.5cm][16cm,29cm]{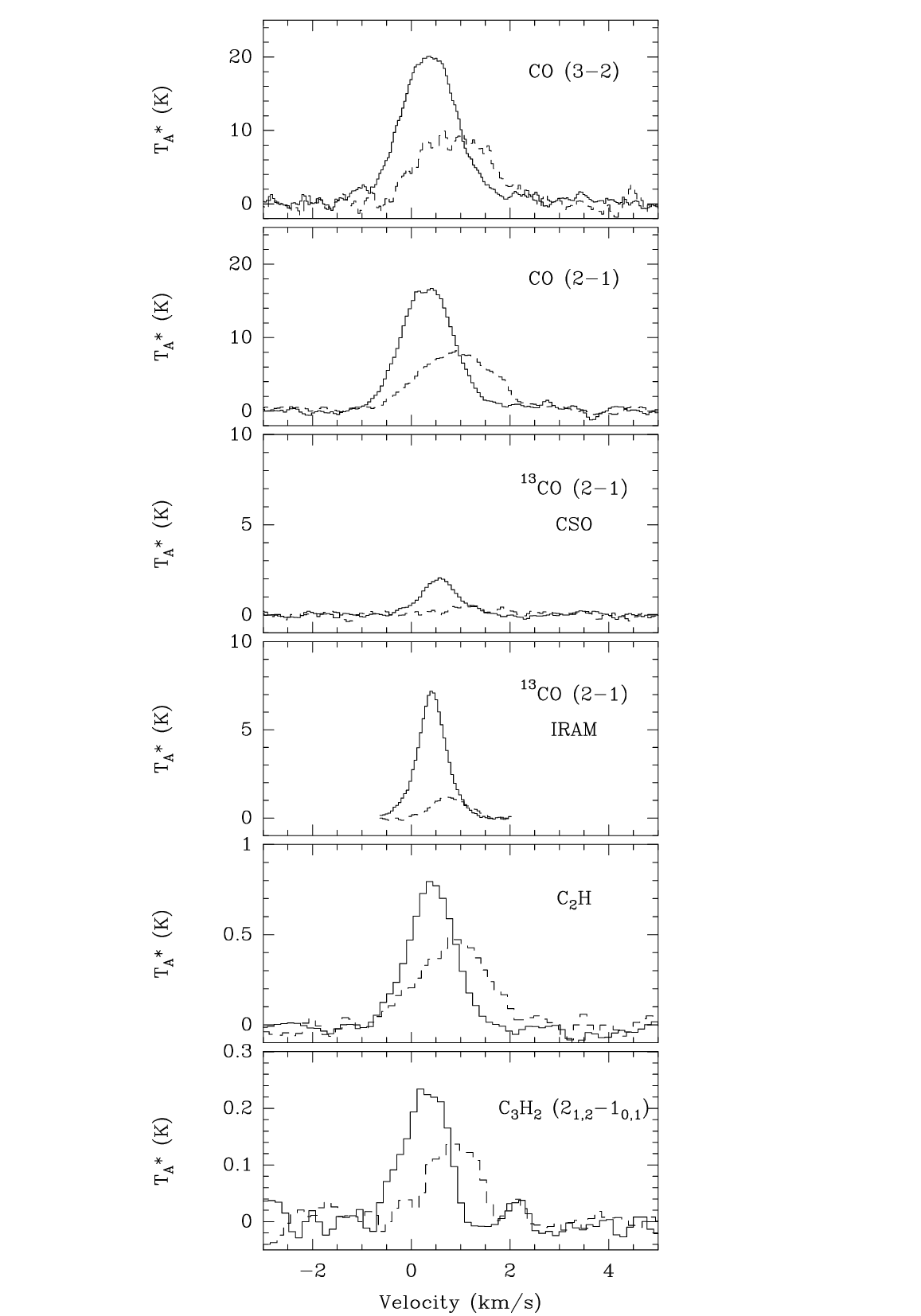}}
\caption{Left : Image of IC~63 from the Digital Sky survey, 
with contours of the $^{12}$CO(2-1) emission. 
Right : comparison of spectra towards the CO peak (full line) 
and the "tail" (dashed line).}
\label{fig:dss}
\end{figure}

\begin{figure}[ht]
\resizebox{\hsize}{!}
{\rotatebox{270}{\includegraphics{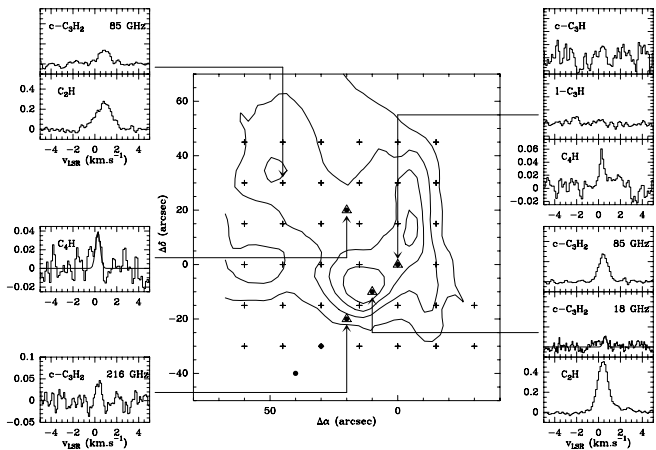}}}
\caption{Example of spectra of various molecules obtained towards 
IC~63, overlayed on a map of the 6.2 $\mu m$ feature obtained with
ISOCAM.}
\label{fig:ic63}
\end{figure}

IC~63 is a small cometary globule illuminated by the B0.5 star 
$\gamma$Cas (HD~5394), at a distance of 230 $\pm 70$ pc. The PDR has
been studied and characterized by \cite*{Jan:94} and 
\cite*{Jan:96}. The gas density is rather high in the "head" of the nebula, 
$5 \times 10^{4}$ cm$^{-3}$ but slightly lower in the "tail" 
(10$^4$ cm$^{-3}$). The gas is warm with T$_K = 50$ K due to the 
intense radiation field, $\chi = 1100$ in units of the Habing  ISRF 
(\cite{Habing:68}). From the overall shape of the nebula, it is likely 
that the molecular gas is currently photo-dissociated and evaporated by 
the intense radiation. Indeed the total column density of molecular
hydrogen is rather moderate for a dense cloud, with a peak value of 
N(H$_2$) = $5\pm 2\times 10^{21}~cm^{-2}$ (\cite{Jan:94}). 
Using the IRAM 30m telescope, we have detected CCH, c-C$_3$H$_2$ 
and C$_4$H in IC~63 and searched for c-C$_3$H and l-C$_3$H. Examples
of spectra are shown in Figures \ref{fig:dss} and \ref{fig:ic63}. 
We found that CCH and c-C$_3$H$_2$ are present everywhere in the
nebula, both at the CO peak and in the tail where $^{13}$CO
disappears. Compared to the line profiles at the peak position, the
lines are a factor of two weaker in the tail, and their centroid is 
redshifted. The same difference is also seen in $^{12}$CO spectra 
while $^{13}$CO is barely seen in the tail (Fig. \ref{fig:ic63}). 
Both the large scale distribution and the similarity of profiles 
with $^{12}$CO indicate that CCH and c-C$_3$H$_2$ are present in the 
external layers of the PDR, where the radiation field destroys 
$^{13}$CO. C$_4$H looks similar to CCH and c-C$_3$H$_2$ though the 
S/N is rather low for this species. We derived column densities of 
N(CCH) $\sim 3 \times 10^{13}$ cm$^{-2}$ and N(c-C$_3$H$_2$) 
$\sim 2.0 \times 10^{12}$ cm$^{-2}$, N(C$_4$H) = $ 3 \times10^{12}$ 
cm$^{-2}$  for this source. The upper limits are N(c-C$_3$H) 
$\leq 1.2 \times 10^{12}$ cm$^{-2}$, N(l-C$_3$H) 
$\leq 5.8 \times 10^{12}$ cm$^{-2}$.

\begin{figure}[h!]
\resizebox{\hsize}{!}
{\includegraphics{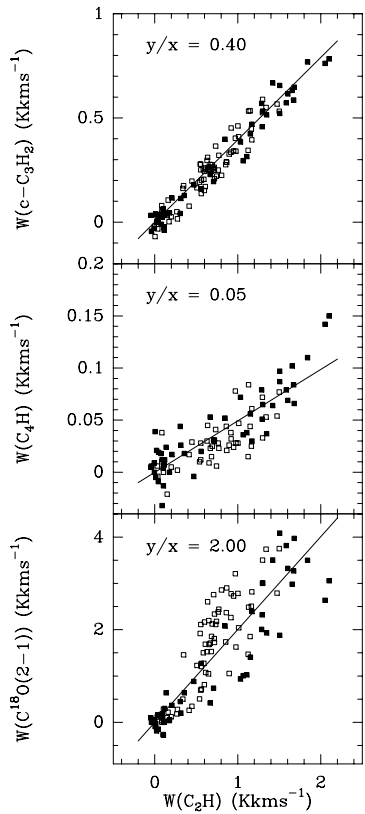}}
\caption{Scatter plot of the integrated emission of CCH (J=1-0
mainline) and c-C$_3$H$_2$ (2$_{12}-1_{01}$ (top), C$_4$H (9-8) 
(middle) and C$^{18}$O (2-1) (bottom) in the horsehead nebula. 
The spatial resolution is 28 arcsec for all lines.}
\label{fig:correl}
\end{figure}

Lines are stronger in the horsehead nebula, where both C$_3$H isomers,
the cyclic and linear forms, are detected. As in IC~63, all
hydrocarbon maps are very similar to each other, but rather different 
from those of CO and isotopes (\cite{Teyssier:02}). A scatter plot
between intensities of CCH, c-C$_3$H$_2$, C$_4$ and C$^{18}$O is shown
in Fig. \ref{fig:correl}. While the correlation is perfect between 
CCH and both C$_4$H and c-C$_3$H$_2$, the scatter plot with C$^{18}$O 
shows a larger dispersion. A closer inspection reveals that the signal 
from CCH remains constants near 1 Kkms$^{-1}$ at the positions the 
furthest away from the cloud edge, while C$^{18}$O lines get brighter
there. From these correlations, and knowing the physical conditions 
in the horsehead nebula, we derived typical abundance ratio for 
the hydrocarbons, which can be compared with the results in diffuse
clouds. They are listed in Table \ref{tab:dif}.  C$_4$H and
c-C$_3$H$_2$ 
have very similar abundances in PDRs, and are about one order of magnitude 
less abundant than CCH. Both C$_3$H isomers are less abundant than 
c-C$_3$H$_2$. The same is true for the linear isomer l-C$_3$H$_2$
compared to the cyclic isomer.

\subsection{The carbon budget}
\vskip 0.3cm
\begin{table}
\begin{center}
\caption{Carbon Budget in PDRs}
\renewcommand{\arraystretch}{1.2}
\begin{tabular}[h]{lrr}
\hline
\hline
Number of Carbon Atoms & $ {\rm N_{Tot}} \over {\rm N(H_2)}$ & 
${\rm N(C)} \over {\rm C_{Total}} $ \\
\\
\hline
1 (C$^+$, C, CO, ...) & $2.6 \times 10^{-4}$ & 1 \\
1 (CH, CH$^+$, ...) & $7.0 \times 10^{-8}$ & $2.7 \times 10^{-4}$ \\
2 (C$_2$, C$_2$H, ...) & $8.0 \times 10^{-8}$ & $6.0 \times 10^{-4}$\\
3 (C$_3$,c-C$_3$H$_2$, ...) & $5.0 \times 10^{-9}$ & $6.0
\times10^{-5}$ \\
4 (C$_4$, C$_4$H, ...) & $2.0 \times 10^{-9}$ & $3.0 \times 10^{-5}$
\\
\hline
\end{tabular}
\label{tab:budget}
\end{center}
{\small \it This table shows for molecules with a given number of carbon atoms 
(Column 1), the total abundance relative to H$_2$ (Column 2) and 
the total number of carbon atoms locked in molecules this size
compared to the available number of Carbon atoms in the gas phase 
(Column 3). We used a gas phase Carbon abundance realtive to H of 
$1.3 \times 10^{-4}$. }
\end{table}

\begin{figure*}[h!]
\centering
\resizebox{\hsize}{!}
{\includegraphics[width=0.8\linewidth]{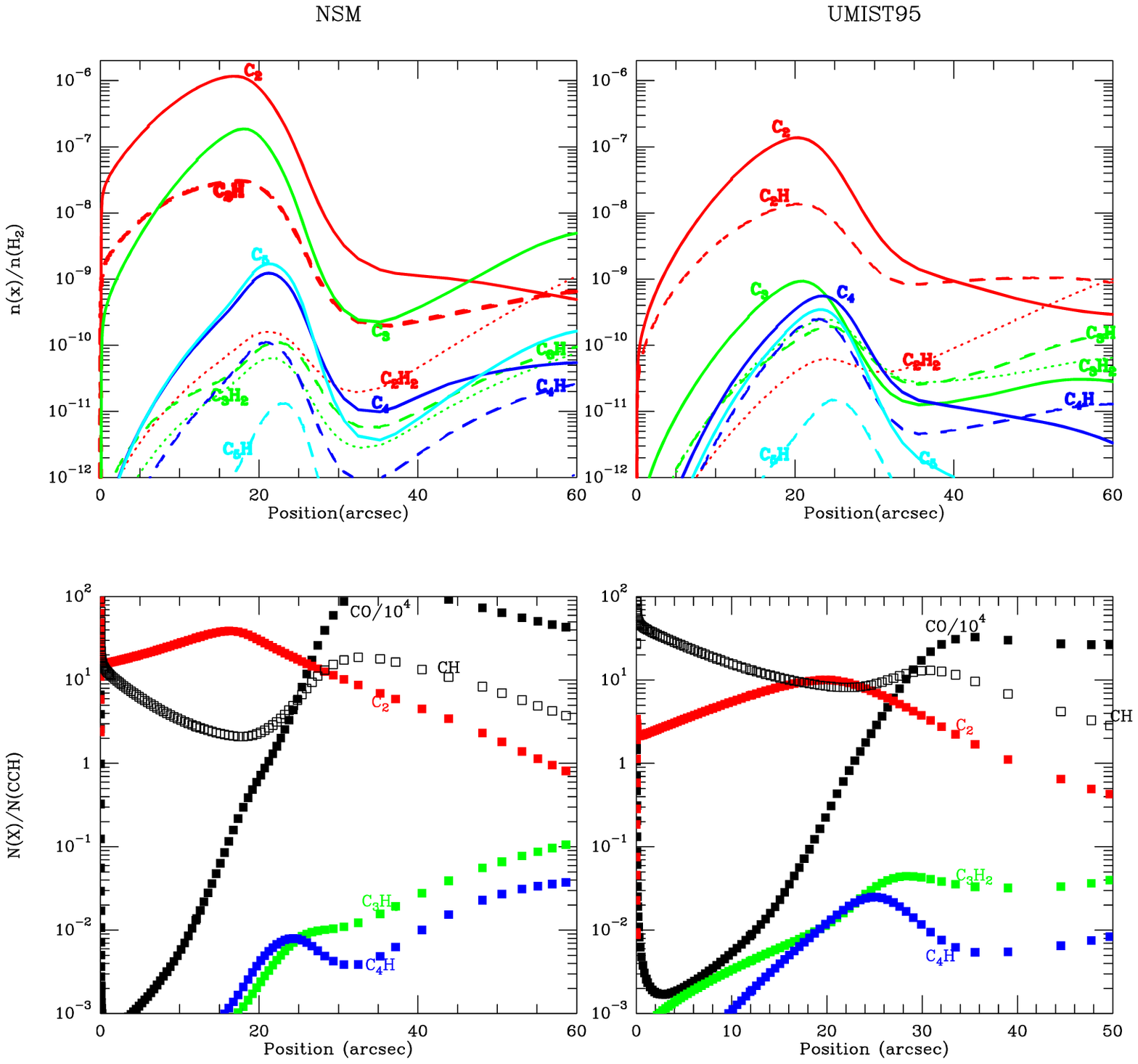}}
\caption{Prediction for the abundances of various hydrocarbons in a 
PDR with a gas density $n=2 \times 10^4$ cm$^{-3}$ and a radiation 
field G$_0$ = 100. Predictions using the NSM reaction set are shown on 
the left side, and using UMIST-95 on the right side. Both models
predict too low abundances for c-C$_3$H$_2$ and C$_4$H, and 
spatial distributions different from the observations. 
However, both models predict that CCH is present in the UV illuminated
gas as observed. }
\label{fig:model}
\end{figure*}

Abundances and abundance ratio of carbon chains are very similar in
PDRs and in diffuse gas. We present below a tentative carbon budget, 
combining the information obtained at all wavelengths in diffuse
clouds and in PDRs. Since all molecules are not observed in the same 
sources, this carbon budget has only a statistical meaning, 
and  numbers cannot be very accurate. We have coadded typical
abundances for molecules with a given number of carbon atoms, 
from one to four (see Table \ref{tab:budget}). Nevertheless, the
comparison of the fraction of carbon atoms locked into 
molecules of various sizes (Column 3) shows a clear trend.
It decreases steeply with the number of carbon atoms for small
molecules, but then tends to flatten for molecules having 
three or more carbon atoms. If the trends continue for larger species, 
the detection of larger carbon clusters or carbon chains looks 
promising for the near future. The only work so far is the negative 
search for C$_4$ and C$_5$ towards $\zeta$ Oph (\cite{Maier:02}). 
Despite the brightness of this star the limits achieved so far do no 
rule out the presence of these carbon clusters with an abundance 10
times lower than C$_3$. Searches are necessary in more opaque lines of 
sight where larger column densities for all species are expected.

\subsection{Models}

Chemical models are now capable of dealing with very large 
reaction rates sets, including complex molecules. We used two 
different reaction rates sets, the UMIST-95 file (\cite{Leteuff}) 
and the New standard model (NSM - \cite{Herbst}) coupled with the 
Meudon PDR code (\cite*{LB:93}, and 
http://aristote.biophy.jussieu.fr/MIS/). 
Results are shown in Figure \ref{fig:model} for physical conditions 
adapted to the horsehead nebula, namely a gas density of 
$n = 2 \times 10^4$ cm$^{-3}$ and a radiation field 100 times more 
intense than the mean value, G$_0$ = 100. We show for each model 
the variations of selected molecular abundances as a function 
of the distance to the cloud edge, as well as the spatial
variations of  abundance ratio relative to CCH. A good spatial
correlation with CCH (as c-C$_3$H$_2$ in Fig. \ref{fig:correl}) would
produce a constant abundance ratio in these plots. 
Predictions obtained from both the 
UMIST-95 and NSM sets follow the same general trend with a peak of 
carbon chains  at $\sim 20$ arcsec (A$_V$ = 2 mag) and a second peak 
deeper into the cloud. Also the chemistry in the UV dominated region 
is very simple, with pure carbon clusters (C$_2$, C$_3$, etc.) 
being more abundant that  radicals (C$_2$H,C$_3$H, etc.) 
which are themselves more abundant than more hydrogenated species 
(C$_2$H$_2$, C$_3$H$_2$, etc.). Finally abundance ratios of both
C$_3$H$_2$ and C$_4$H with CCH do not vary much inside the cloud
(for Positions larger than 20 arcsec) but decrease sharply at the
cloud edge.  This simple behavior is in contrast 
with our measurements :\\
-- C$_3$H$_2$ is always more abundant than C$_3$H, this is true 
for the cyclic isomer alone and when different isomers are combined.\\
-- C$_4$H and c-C$_3$H$_2$ have almost the same abundance, 
$\sim 2 \times10^{-9}$, a factor of ten at least larger than the 
predicted value\\
-- All hydrocarbons observed are detected in the UV illuminated region 
where the PAH emission is observed, while the model predict a spatial 
offset. 

While models are able to reproduce fairly well the formation of CO,
and marginally the CCH abundance, severe problems remain for 
larger molecules. Since 
the main destruction process is photodissociation -- an efficient
mechanism in PDRs -- new formation routes must be found for explaining 
the large abundances of C$_4$H and C$_3$H$_2$ observed in PDRs. 
It is likely that some reaction rates used in current models are 
inaccurate, and that important reactions are still missing. 
We need more observations on simple species to progress, especially 
on neutral and ionized carbon which are the main reactants in the 
chemical reaction networks.

\section{Chemistry in dark clouds}

\subsection{Carbon chemistry}
Since the same networks are used to probe dark cloud and PDR
chemistry, we may learn on the formation processes by comparing
abundances and abundances ratio in different environments. 
TMC-1 in the Taurus molecular cloud is known as the ISM "carbon
factory", in which long carbon chains and cyanopolyyne have been
detected.
Neutral carbon itself is known to be abundant both in 
TMC-1 (\cite{Schilke:95}, 
\cite{Maezawa:99} and in L~134N (\cite{Tatematsu:99}). According to 
these authors, the constancy of the antenna temperature of the ground 
state fine structure line of atomic carbon in dark clouds, 
T$_A^*$ $\sim {\rm 2 K}$, indicates that atomic carbon is abundant in the 
sources and acts as a coolant. The corresponding kinetic 
temperature is about 9 K, a typical figure for dark clouds. 
Since the emission is very extended, it is likely that carbon is
present in the molecular envelope, and helps in cooling this envelope 
down to 10 K. Is the presence of long carbon chains
 related to neutral carbon ? 
 Are long carbon chains  present in other dark clouds ?
We performed a restricted survey of carbon chains in L134N, to 
have a comparison with TMC~1 and answer these questions.
The results will be described in more detail elsewhere
(\cite{Fosse:03}), and are just summarized below. 

In deep integrations we obtained new detections of 
C$_3$H (cyclic and linear), C$_4$H and C$_6$H in L134N. 
A CCH map is shown in \cite*{Dickens:00}. Compared to 
molecules tracing the cloud core  such as NH$_3$ and N$_2$H$^+$ 
(e.g. \cite{Pagani:03},  \cite{Bergin:03}), 
CCH peaks are spatially separated from dense gas but coincide with
those of C$_4$H and other hydrocarbons (\cite{Dickens:00},
\cite{Fosse:03}). From the low line intensity, and 
the different spatial distribution, we conclude that carbon chains 
reside in cold but not very dense molecular gas in L134N: 
n(H$_2$) = $3 \times 10^3$ cm$^{-3}$ and T = 10 K. 
The column densities are nevertheless rather large, and
reach similar values as towards the cyanopolyyne peak in TMC-1, 
namely N(CCH) $\sim 1.5 \times 10^{14}$ cm$^{-2}$, 
N(C$_4$H) = $\sim 1.3 \times 10^{14}$ cm$^{-2}$ and 
N(C$_6$H) =  $\sim 1.2 \times 10^{12}$  cm$^{-2}$. 

These conclusions are consistent with the presence of atomic carbon 
in dark cloud envelopes. Carbon chains, which are detected both in
diffuse or dark clouds, and also in PDRs,  seem to be always seen
in places with rather large abundances of carbon, either atomic or
ionized. Though unexpected, since the formation of hydrocarbons is
driven by chemical reactions involving C and C$^+$, the observed abundances
and abundance ratio constrain efficiently current models.
 
\subsection{Carbon budget in dark clouds}

\begin{table}
\begin{center}
\caption{Carbon Budget in Dark Clouds}
\renewcommand{\arraystretch}{1.2}
\begin{tabular}[h]{lcc}
\hline
\hline
Number of Carbon Atoms & $ {\rm N(Tot)} \over {\rm N(H_2)}$ & 
$ {\rm N(C)} \over {\rm C_{Total}}$ \\
\\
\hline
1 (C$^+$, C, CO, ...) & $8 \times 10^{-5}$ & 1  \\
1 (CH, HCN, CS ...) & $\leq 1.0 \times 10^{-7}$ & $\leq 1.2 \times 10^{-3}$ \\
2 (C$_2$H, C$_2$S...) & $1.0 \times 10^{-8}$ & $2.5 \times 10^{-4}$\\
3 (c-C$_3$H$_2$, HC$_3$N, CH$_3$CCH ...) & $1.2 \times 10^{-8}$ & $4.5 
\times 10^{-4}$ \\
4 (C$_4$H, C$_4$H$_2$, ...) & $1.7 \times 10^{-8}$ & $8.5 \times 10^{-4}$ \\
5 (HC$_5$N, C$_5$H, ...) & $1.2 \times 10^{-9}$ & $7.5 \times 10^{-5}$ \\
6 (C$_6$H, C$_6$H$_2$...) & $4.2 \times 10^{-10}$ & $3.1 \times 10^{-5}$ \\
7 (HC$_7$N, ...) & $5.5 \times 10^{-10}$ & $4.8 \times 10^{-5}$ \\
8 (C$_8$H, ...) & $1.1 \times 10^{-11}$ & $1.1 \times 10^{-6}$ \\
9 (HC$_9$N, ...) & $1.2 \times 10^{-10}$ & $1.3 \times 10^{-5}$ \\
11 (HC$_{11}$N, ...) & $1.4 \times 10^{-11}$ & $1.9 \times 10^{-6}$ \\
\hline
\end{tabular}
\label{tab:budgetbis}
\end{center}
{\small \it
The table shows for molecules with a given number of carbon atoms, the total 
abundance relative to H$_2$ in TMC-1 and the total number of carbon atoms 
compared to the available number of Carbon atoms in the gas phase. 
We assumed a gas phase CO abundance relative to H$_2$ of 
$8.0 \times 10^{-5}$ and a total H$_2$ column density of 
$2 \times 10^{22}$ cm$^{-2}$. Data on TMC-1 from \cite*{Tur:00}, 
\cite*{Pratap:97}, 
\cite*{Bell:98}, \cite*{Fosse:01}, \cite*{Dickens:01}.}
\end{table}

As for  PDRs we have computed the carbon budget for cold
dark clouds. Due to the presence of ice mantles on dust grains, 
the absolute column densities of H$_2$ are uncertain in this case. 
Nevertheless the numbers quoted
in Table \ref{tab:budgetbis} can be compared, at least relative to each
other. We used TMC-1 as most molecules have been detected towards
this source, and assume a total extinction of 30 magnitudes, a
column density of N(H$_2$) = $2 \times 10^{22}$ cm$^{-2}$ and a
gas phase CO abundance relative to molecular hydrogen of $8.0 \times
10^{-5}$. Depletion on ices is causing a deficit of gas phase carbon
compared to the situation in diffuse clouds, as
seen  by the lower abundance of gas phase CO relative
to H$_2$.
In addition to CO, gas phase carbon is locked in organic molecules 
with 3 or 4 carbon atoms in TMC-1 (cf Table \ref{tab:budgetbis}). 
There is a clear decrease of the fraction of carbon tied in 
 more complex molecules.  It is likely that the available information
on organic molecules with 5 or more carbon atoms is 
incomplete,  since a lower number of molecules this size
are detected, compared with smaller molecules.
Other species might still remain 
to be identified. For example, pure carbon
clusters are likely present in dark clouds, but could not been taken into
account in the above statistic due to the lack of definite detections.

\begin{figure}[h]
\centering
\resizebox{\hsize}{!}{\includegraphics*[0.5cm,7.0cm][20.0cm,26cm]{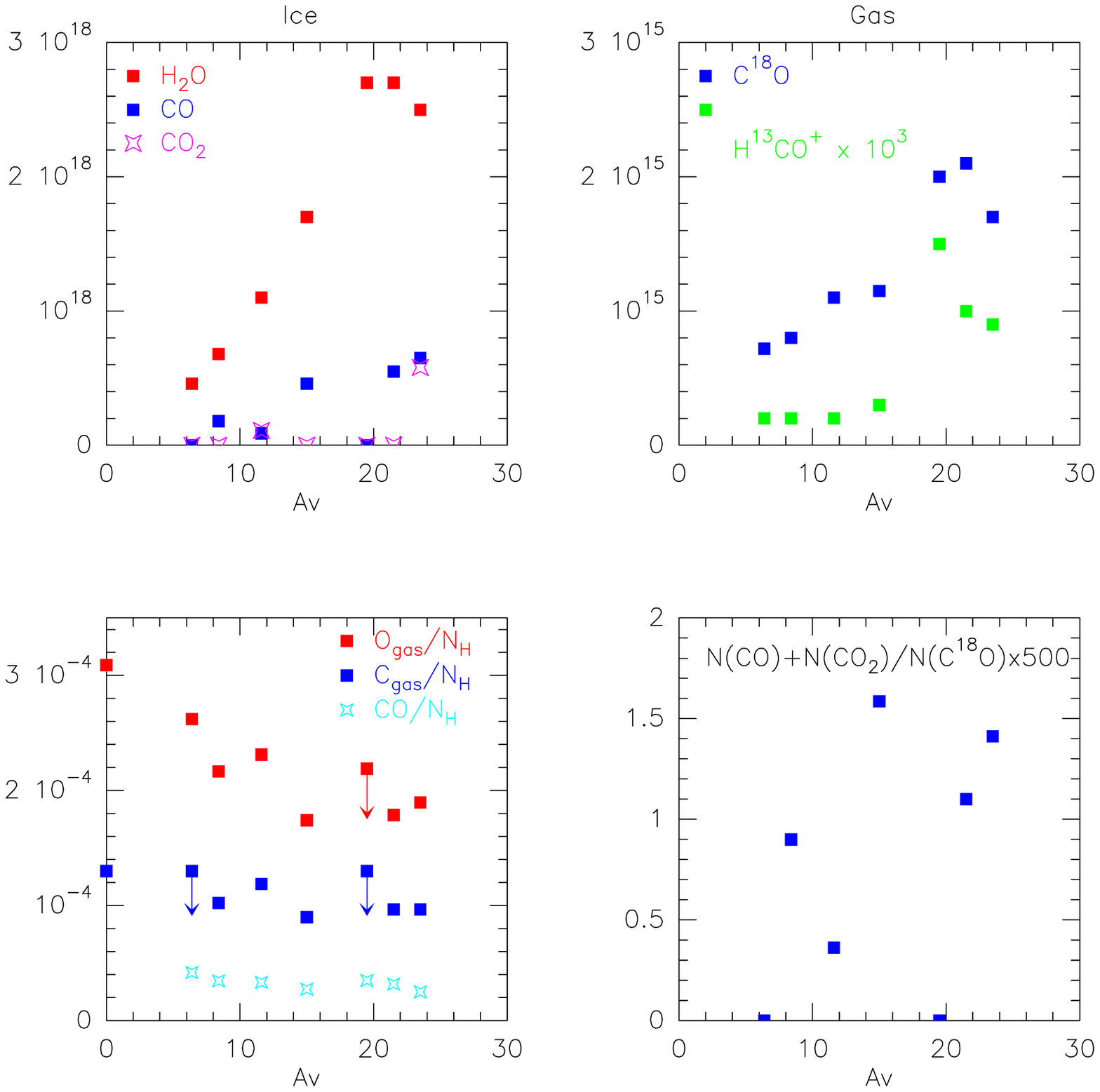}}
\caption{Column densities of molecular ices and gas phase species 
for background sources in the Taurus Molecular Cloud. We used 
$N_H (cm^{-2}) = 1.52 \times 10^{21} \times A_V (mag)$, 
$C_{gas} = {{1.3 \times 10^{-4} (N_H - N_{ice}(CO) - N_{ice}(CO_2)) } 
\over {N_H}}$ and $O_{gas} = {{3.09 \times 10^{-4} (N_H -
N_{ice}(H_2O) - N_{ice}(CO)  - 2N_{ice}(CO_2)) } \over {N_H} }$. 
We assume $N_{ice}(CO_2) = N_{ice}(CO)$ for the lines of sight with no 
CO$_2$ observations. Data from Teixeira and Emerson (1999), 
Nummelin et al. (2001) and Murakawa et al. (2000). }
\label{fig:depletion}
\end{figure}

\subsection{Chemistry and depletions}

The widespread presence of molecular ices on dust grains in 
shielded regions is now well established. These ices contribute in 
modifying the grain optical properties and in favoring grain
coagulation. In the Taurus cloud, water ice is seen for lines of sight 
with extinction larger than a threshold of A$_V$ $\sim 3$ mag,
\cite*{Teixeira:99}, while CO ice appears for larger A$_V$s, 
near 6 magnitudes. It is therefore certain than many
gas phase molecules, including carbon monoxide and water, are condensed 
on grain mantles in cloud interiors, though a few of them only
have an solid phase abundance high enough to be detectable in
the infrared (the typical limit is a few \% of the water ice, or a few
times 10$^{-6}$ relative to hydrogen, \cite*{nummelin:01}).
 The inventory of ices has
been completed thanks to the ISO mission, both towards YSOs and
towards background field stars. These lines of sight are particularly 
interesting for chemistry since they sample random lines of sight
across clouds, which are likely representative of the bulk of 
shielded molecular gas. Both the total extinction and the chemical 
composition of the solid phase are fairly well known.

 We have selected a sample of seven lines of sight in the 
Taurus molecular clouds -- two of them  being located in TMC-1 -- 
to investigate the relationship between solid and gas phase 
chemical composition. We used the IRAM 30m telescope to perform 
the observations in December, 2000.  Properties of the background
sources are summarized in Table \ref{tab:depletion}.
Fig. \ref{fig:depletion} shows
that, while CO isotopes are readily 
detected for all lines of sight with A$_V$ between 6 and 21
magnitudes, molecular ion lines are weak or absent for A$_V$ lower
than $\sim$ 15 magnitudes. At this point, 
the amount of carbon in the solid phase (condensed as CO or CO$_2$
ice) is larger than the amount of carbon in the gas phase 
(primarily as CO). Dense prestellar cores have similar 
C$^{18}$O column densities as the lines of sight with the largest
extinction in our sample, $2 \times 10^{15}$ cm$^{-2}$ for A$_V$ = 20
mag but larger column densities of molecular ions. Because the total
extinction through these dense cores is very large, at least 
50 mag or even more are derived from star counts in L134N
(\cite{Pagani:03}), it is expected that the remaining fraction
of molecules in the gas phase is low. Indeed, 
indirect evidences of large depletions of 
molecules onto grains have been given based on comparisons 
with the the continuum emission from dust grains, star counts, 
and the spatial distribution of different molecules (see e.g. 
\cite{Pagani:03};  \cite{Bergin:03}), in agreement with the moderate
depletions reported here for lower extinctions. 
Note that because CO gets converted into CO$_2$ and possibly 
other species on grains, the amount of carbon condensed on grains, 
hence the impact of molecular ices on the chemistry, is larger 
than what can be derived from the concentration of CO ice only. An
inventory of the ice composition, including CO$_2$ and possibly
CH$_3$OH, is necessary for determining the carbon and oxygen budget
of the solid phase.

It is interesting to analyze the chemical differences in 
TMC-1 itself. Two lines of sight through TMC-1 belong to our sample, 
Tamura~8  in the ammonia rich region (North) and E~0484 in the carbon 
rich region (South) and have similar total extinction of $\sim 20$
mag and water ice column densities $2.7 \times 10^{18}$ cm$^{-2}$,
 while the gas 
phase abundances we observed follow the general trends in TMC-1 of 
larger column densities of molecular ions in the ammonia rich region 
compared to the carbon rich region.  Since the formation of molecular 
ices in clouds is a rather slow process (\cite{Ruffle:0}, 
\cite{Ruffle:1}), the presence of molecular ices everywhere in the 
Taurus cloud, with well defined dependence on total extinction but no 
dependence on the gas phase composition, is a strong 
argument against different ages for different regions. 
For most of the lines of sight, the abundance of water ice relative to
molecular hydrogen (deduced from A$_V$) is comparable or larger than
$ 10^{-4}$, which means that a significant fraction of the
available oxygen is condensed. 
Other  processes such as those discussed below are more likely 
to contribute in producing the marked gas phase chemical differences. 
A deeper investigation of the selected lines of sight in TMC-1 may
bring clues on the origin of the chemical differentiation since both 
the total gas column density and the composition of dust grains are known.

\begin{table}
\begin{center}
\caption{Background stars  in the Taurus  molecular cloud }
\renewcommand{\arraystretch}{1.2}
\begin{tabular}[h]{lccccc}
\hline
\hline
Name & A$_V$  & H$_2$O$^*$ & CO$^*$ & CO$_2^*$ & CO \\
& mag &  &  &  &   \\
\hline
Elias~6 & 6.4 & 4.6  & ... & ... & 3.6  \\
Elias~3 & 8.4 & 6.8  & 1.8  & ... & 4.0 \\
Elias~13 & 11.6 & 11  & 0.9 & 1.1 & 5.5  \\
Elias~15 & 15.0 & 17  & 4.6  & ... & 5.8 \\
E~04384 & 19.5 & 27  & ... & ... &  10 \\
Tamura~8 & 21.5 & 27  & 5.5  & ... &  10 \\
Elias~16 & 23.5 & 25 & 6.5  & 8.5 \\
\hline
\end{tabular}
\label{tab:depletion}
\end{center}
{\small \it For sources listed in column (1), Column (2) gives the
extincion, Column (3) the column density of water ice, Column (4) the
column density of CO ice, Column (6) the column
denisty of CO$_2$ ice, and Column (6) the column density of gas
phase CO deduced from C$^{18}$O data (see text). All column
densities are given  in units of 10$^{17}$ cm$^{-2}$. Data from
Teixeira and Emerson (1999), Nummelin et al. (2001) and 
Murakawa et al. (2000).}
\end{table}

\section{Discussion and Perspectives}

In all interstellar environments, dark and diffuse clouds, and PDRs,
 carbon chains and cycles are closely related to each
other, and to atomic carbon. They are present in large abundances 
in the moderately dense molecular gas but avoid the darkest and 
densest cores. This conclusion is valid for the pure carbon chains 
and cycles. Other carbon species, like cyanopolyynes are found 
in high density regions. Though their formation mechanism also 
involves carbon, the difference between nitrogen and carbon chemistry 
may contribute to the different spatial distribution.
The difference of incident radiation between diffuse or UV illuminated gas on
one side, and dark clouds on the other side, affects somewhat the relative
abundance of carbon molecules in permitting the growth of larger
chains (compare the abundance of C$_4$H in TMC-1 and in the horsehead
nebula) but the chemistry looks nevertheless very similar.
Another major difference between diffuse or UV illuminated gas on
one side, and dark clouds on the other side, is the condensation
of abundant gas phase molecules on dust grains. The main impact of 
condensation on the chemistry is the enhanced abundance of molecular
ions in depleted region. The reason for this is that the destruction channels
for molecular ions are not only dissociative recombination reactions
with electrons, but also reactions with abundant gas phase molecules. 
The depletion of CO on dust grains therefore leads to lower
destruction rates for molecular ions and larger abundances.
While the composition of molecular ice depends mainly on the
total extinction, marked chemical differences are found
in lines of sight with similar column densities.
Also, measured abundances of carbon chains and cycles
are often larger than steady state predictions.
The usual explanation for understanding the large abundances, is to invoke
the so-called "early time chemistry", in which clouds are seen in
an early evolutionary stage since their formation, in which the 
available carbon, initially ionized,  has not be completely
transformed into CO. However,
several arguments show  that  "early time chemistry" cannot be
the sole explanation.  Early time chemistry is efficient 
because the formation of molecular 
hydrogen and the conversion of carbon to CO takes a very long time 
($\geq 10^6$ years) when starting from atomic gas. It is thus
tempting to use molecular abundances in dark clouds as a chemical clock.
However the same arguments leading to doubt on ``early time ''
chemistry, are also against this ```chemical clock '' idea
 for dense cores.

i) Since the same molecules, with similar abundances, are seen in 
many different clouds, at different evolutionary stages, such as PDRs, 
pre-stellar clouds, clouds having already formed stars, how could
we be viewing ``early time'' chemistry in all cases ? For example, 
clouds having already formed stars and clouds associated with PDRs 
cannot be presented as ``young'' since they have already had time 
to collapse and form stars, and for PDRs the newborn stars have had
time to disrupt their parent clouds. Even if the evolution is not
uniform across a cloud (chemical time scales depend on the
density, so that more diffuse regions evolve at a slow pace compared
to dense cores) the requirement of ``early time'', i.e. a cloud age lower 
than  10$^5$ years is very strong compared to the typical time
scale for cloud formation and evolution towards 
forming stars ($\geq 10^6$ years, \cite{hartmann:01}). 

ii) A second argument against ``early time'' chemistry is the heavy 
dependence of the predicted chemical abundances on the initial conditions 
\cite*{Lee:96} while observations show well defined chemical properties.

iii) Finally, as stated above, molecular ices are
commonly detected in dark clouds which indicate the gas has had
time to evolve, and built gas phase and solid phase molecules
 (\cite{Roberts:02}).

Other processes are known to strongly affect the chemistry, and may 
 favor the presence of carbon and carbon chains in cloud envelopes. 
The propagation of UV radiation depends on the grain optical
properties. \cite*{Casu:01} have shown that grain growth in dark
clouds affects the grain optical properties hence the propagation of 
UV radiations and the photodissociation rates in dense gas. This
effect leads to higher abundances of neutral and ionized carbon 
compared to models using the grain properties valid for diffuse gas. 
It must be remembered that photons play a very important role in 
interstellar chemistry, even in dark clouds since the 
interaction of cosmic rays with hydrogen leads to the production of UV photons 
deep into the clouds. In addition, the chemical equilibrium for 
moderately dense gas  can be carbon rich, the HIP solution 
(\cite{LB:95}) even without any incident UV photons.
These last two explanations rely on microscopic processes and choice 
of elemental abundances. But the structure and dynamical state of 
interstellar clouds also affects the chemistry. Clumpy
cloud models, such as those of \cite*{spaans:97} or \cite*{storzer:96} 
can explain the widespread distribution of carbon by the
accumulation of many clump surfaces, rich in carbon, in
telescope beams. Since UV radiation can penetrate deeper into
clumpy clouds than in homogeneous structures, each clump surface
behaves as a mini-PDR in this context. Another idea is
to include the effect of turbulence in homogeneous clouds.
The main effect of turbulence is to increase the  diffusion
between the cloud surface and its interior.
 A small level of turbulent diffusion, 
compatible with the observations, can maintain higher abundances 
of atomic carbon and atomic hydrogen in cloud envelopes than 
in steady state models, in agreement with observations of atomic
hydrogen in dark clouds (\cite{Willacy:02}).

It is likely that all additional physical mechanisms described above, both 
microscopic (change of optical properties of dust grains, abundances)
and macroscopic (turbulence, chemistry) and possibly others still to uncover, 
contribute significantly to the chemistry. There are  so many
unknowns in the physical properties and chemistry of dark clouds, 
that deriving the age of clouds from their chemical abundances is not 
reliable. Clues from the age of embedded stars, if any, the geometry, 
the environment etc.  must be used to retrace the history of
interstellar structures. To progress, the whole evolution of
interstellar clouds, from  their formation
 in the warm neutral gas, to their concentration in molecular gas, 
and finally the birth of new stars, must be better understood than
it is now.

\begin{acknowledgements}
The authors thank J. Cernicharo for letting them use his radiative 
transfer code,  D. Teyssier and A. Abergel for using their work 
on the horsehead nebula. The authors thank J. Pety for helping 
with the processing of IRAM plateau de Bure data.
\end{acknowledgements}

\end{document}